\documentclass[prl,twocolumn]{revtex4}

\usepackage[dvips]{graphicx}
\usepackage{amssymb}
\usepackage{amsmath}

\begin{document}

\title{Friction as a probe of surface properties of a polymer glass}

\author{Lionel Bureau}
 \email{bureau@insp.jussieu.fr}
\affiliation{Institut des Nanosciences de Paris, UMR 7588 CNRS-Universit\'e Paris 6, 140 rue de Lourmel, 75015 Paris, France}


\begin{abstract}
We probe the temperature dependence of friction at the interface between a glassy poly(methylmethacrylate) lens and a flat substrate coated with a methyl-terminated self-assembled monolayer. The
monolayer exhibits density defects which act as pinning sites for the polymer chains. We show that the shear response of such an interface supports the existence, at the surface of the glassy polymer, 
of a nanometer-thick layer of mobile chains. Friction can be ascribed to the interplay between viscouslike dissipation in this layer and depinning of chains adsorbed on the substrate. We further show that the
pinning dynamics is controlled by $\beta$ rotational motions localized at the interface.
\end{abstract}

\maketitle

\section{Introduction}
\label{sec:intro}

Polymeric materials are encountered in a variety of situations in which their tribological behavior is of importance, from sliding of rubber components \cite{grosch,chaudhury,bo1} to lubrication via 
surface modification using polymer thin films \cite{BT,klein,donnet}. The fundamental understanding of polymer friction has been the scope of a number of studies over the past fifty years, from which the following two main pictures emerge.

Rubber friction involves the combination of an interfacial molecular process and of bulk viscoelastic losses, the respective weight of these two mechanisms depending on the
roughness of the substrate on which the elastomer slides \cite{grosch}. Schallamach described the interfacial process as the thermally activated formation and breaking, under the applied shear stress, of 
molecular bonds between the rubber and the countersurface \cite{schal}. Such a pinning/depinning mechanism, which serves as a basis for more refined models \cite{chern,klaft}, 
is consistent with the velocity-dependent friction reported 
in experimental studies of rubbers sliding on smooth substrates \cite{grosch,chaudhury}.

Friction of glassy polymers is usually related to plasticity \cite{BT,thomas,briscoe2,epje1}. In the absence of ploughing (or indentation) of the material, {\it i.e.} in the absence of bulk dissipation, friction is 
found to be controlled 
by the yield properties of a nanometer-thick polymer layer confined at the interface between the solids \cite{epje1,epje2,revue,tirrell1}. 

Now, the issue of surface properties of glassy polymers has recently attracted much attention \cite{keddie,forrest1,GBM}. Various techniques have been used for this purpose:  glass transition 
temperature measurements on ultra-thin
films \cite{forrest1,nealey,ellison}, density gradient determination \cite{algers}, nanoscale contact mechanics \cite{forrest2}, creep \cite{McK} and buckling \cite{stafford} of thin films , or 
friction force microscopy \cite{hamm,sills,sills2}. 
Though some of the results are still a matter of debate, there is a consensus that a glassy polymer exhibits a nanometer-thick surface layer in which the chain mobility differs from that of the bulk. 
Moreover, the dynamics in this layer is 
strongly influenced by the nature of the chemical interactions between the polymer and the substrate: mobility appears to be enhanced near a free polymer surface or an interface with  a low energy substrate,
whereas chain dynamics seems to be  slowed down in the case of strong polymer/substrate interactions \cite{nealey,torres,varnik}.

In this context, we have recently shown that such a dependence of surface chain mobility on interactions has a clear signature on friction \cite{prl97}. We have performed friction experiments in which a
poly(methylmethacrylate) (PMMA) solid slides on smooth surfaces presenting different densities of pinning sites available for polymer/substrate bond formation. We have found that: 

(i) at high pinning level,
frictional dissipation occurs through the sudden flips of molecular-sized bistable regions localized in a nm-thick layer of confined chains, which responds to shear as an elasto-plastic solid \cite{bo,revue}, 
and 

(ii) in situations of weak pinning, dissipation appears to be governed by a process akin to that proposed for rubber friction.

This suggests that some ``glass-to-rubber'' transition occurs at the polymer surface when its interaction with the substrate goes from strong to weak. 

In the present paper, we investigate further the regime of weak pinning
by probing the temperature dependence of friction at an interface between PMMA and a silicon wafer grafted with an organic self-assembled monolayer (SAM). The contact configuration of our experiments, in
which a smooth lens of polymer is pressed, under low contact pressure, on a rigid flat substrate, allows us to probe the shear response of the polymer 
surface without inducing bulk dissipation during sliding. Our results provide strong support for the presence of a 
nanometer-thick layer at the polymer surface, which, in the case of weak interaction with the countersurface, exhibits a frictional rubberlike response. 
This response results from the combination of viscous flow in this surface layer, and of
a pinning/depinning mechanism. Moreover, we show that the pinning dynamics of the polymer chains is controlled by localized $\beta$ rotational motions at the interface.

\section{Experimental section}
\label{sec:exp}

\subsection{Setup}
\label{subsec:setup}

\begin{figure}[htbp]
$$
\includegraphics[width=8cm]{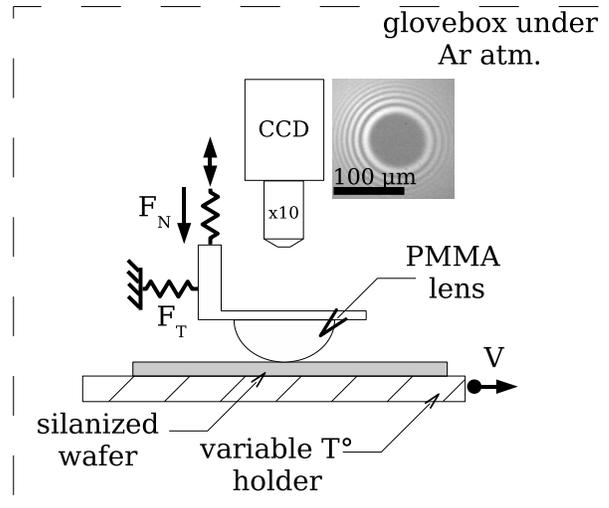}
$$
\caption{Experimental setup: a PMMA lens is pressed against a silanized substrate under a constant normal load. The substrate is mounted on a temperature-controled stage, and is driven at velocity $V$. 
The contact area is monitored optically.}
\label{fig:setup}
\end{figure}

The experiments were performed using a home-built tribometer which is sketched on Fig. \ref{fig:setup}. A lens of poly(methylmethacrylate) is fixed on a transparent holder attached to a load cell made of two 
double cantilever springs of stiffness $K_{N}=2.10^4$ N.m$^{-1}$ and $K_{T}=1.7\times10^{4}$ N.m$^{-1}$. The load cell is mounted on a vertical motorized translation stage which is used to 
bring the lens in contact with a flat horizontal substrate. The value of the applied normal force, $F_{N}$, is deduced from the spring deflection, measured by means of a capacitive displacement gauge. 
The range of accessible normal forces $F_{N}$ is 4.10$^{-3}$---3 N. The silicon wafer used as a substrate is fixed to a horizontal translation stage and is moved at constant velocity $V$ in 
the range 3.10$^{-2}$---10$^{2}$ $\mu$m.s$^{-1}$. The resulting tangential force $F_{T}$ is measured to within 10$^{-3}$ N by means of a capacitive sensor. 
In order to work at constant normal load
during sliding, and to compensate 
for parallelism defects of the mechanical setup, a digital feedback loop controls the position of the vertical stage which drives the loading spring. 

The silicon substrate is mounted on the top plate of a
 heating/cooling unit made
of a thermoelectric element and a water-circulating heat exchanger. The temperature of the plate is measured by means of a thermistor glued on its surface, and is controlled to within 0.1$^{\circ}$C in the 
range from
$-20^{\circ}$C to $100^{\circ}$C. In order to avoid moisture or ice formation at low temperature, the whole experimental setup is enclosed in a glovebox purged with dry argon, which allows us to keep the 
relative humidity below 5\%.

The contact area $A$ between the 
polymer lens and the wafer is monitored in reflection by means of a long working distance optics and a computer-interfaced CCD camera. 
The lenses have radii of curvature on the order of a millimeter (see section \ref{subsec:samples} below), which --- for $F_{N}$ in the range given above --- 
yields contact areas typically ranging from 3.$10^{2}$ to 3.$10^{4}$ $\mu$m$^{2}$. Contact areas are determined with a $\pm 2\%$ accuracy by image processing.

Note that the elastic contrast between the polymer lens (Young modulus $E\simeq 3$ GPa) and the substrate ($E\simeq 100$ GPa), along with the smoothness of both surfaces in contact (see below), ensures
that no ploughing of the polymer occurs during sliding.

\subsection{Samples}
\label{subsec:samples}

\begin{figure}[htbp]
$$
\includegraphics[width=8cm]{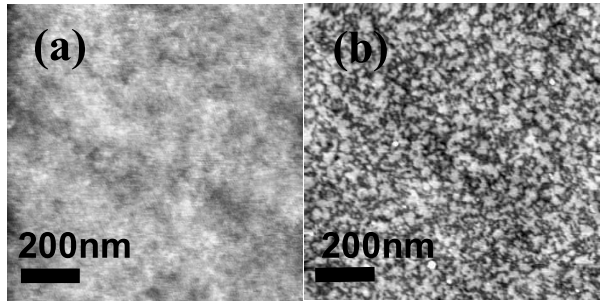}
$$
\caption{AFM topographic images (tapping mode) of (a) the surface of a PMMA lens, and (b) the OTS layer grafted on a silicon wafer. Height scale is 28\AA~from black
(low) to white (high).}
\label{fig:AFM}
\end{figure}

Polymer lenses are made as follows: about 10 mm$^{3}$ of poly(methylmethacrylate) (PMMA) powder (${M_{w}}$=93 kg.mol$^{-1}$, $M_{n}=46$
 kg.mol$^{-1}$, $T_{g}\simeq 110^{\circ}$C,
 from Sigma-Aldrich) 
 is brought to
 T=200$^{\circ}$C at p=10$^{-1}$ mbar until a clear and homogeneous melt is obtained. The melt is then transferred on a clean glass slide and allowed to spread at 
 T=180$^{\circ}$C and atmospheric pressure. During the first minutes of spreading, the highly viscous polymer melt forms a spherical cap, which radius of
 curvature increases with the spreading time. Once the spherical cap has reached a roughly millimetric radius of curvature, it is cooled to 
 80$^{\circ}$C (at $\sim 2^{\circ}$C.mn$^{-1}$) and left at this annealing temperature for 12 hours. The root-mean-square roughness of the lenses at their apex is found to be of 2-3 \AA, as measured
 by atomic force microscopy over a $1\, \mu$m$^{2}$ scan (see Fig. \ref{fig:AFM}a).

The substrate is a 2'' silicon wafer covered by an alkylsilane layer. 
The wafer is first cleaned as follows: rinsing with toluene, drying in nitrogen flux, 
15 minutes of sonication in a dilute solution of detergent in deionized water, 15 minutes of sonication in ultra-pure water, drying in nitrogen flux, 30 minutes in 
a UV/O$_{3}$ chamber. We then graft a self-assembled monolayer (SAM) of octadecyltrichlorosilane (OTS, Sigma-Aldrich), following a procedure akin to that described by 
Silberzan {\it et al.}\cite{silberzan} 
and Davidovits {\it et al.}\cite{goldmann}: the wafer is exposed 
 to a flux of humid oxygen for 2 minutes immediately after UV/O$_{3}$ exposure, and is then immersed in a solution composed of 70 ml of hexadecane, 15 ml of carbon tetrachloride, 
 200 $\mu$l of OTS. It is left for 5 minutes in this reaction bath at 18$^{\circ}$C, then rinsed with carbon tetrachloride. All the reagents are anhydrous grade 
 (Sigma-Aldrich) and used as received. The reaction is conducted in a glovebag under dry nitrogen. Under such conditions, the thickness of the OTS layer, measured by 
 ellipsometry, is 21$\pm 1$ \AA. Atomic force microscopy (Fig. \ref{fig:AFM}b) reveals that the SAM is actually formed of islands of high areal density of OTS molecules (approximately 10--50 nm in size), 
 separated by regions of
much lower coverage density or even bare substrate (also tens of nm in size).
 
\section{Results}
\label{sec:res}

\begin{figure}[htbp]
$$
\includegraphics[width=8cm]{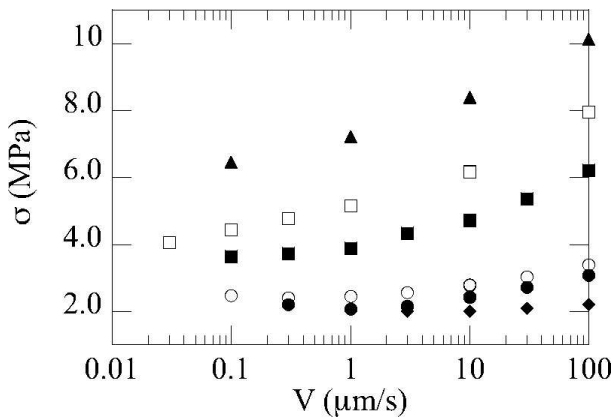}
$$
\caption{Shear stress as a function of velocity, for $p=25$ MPa. ($\blacklozenge$) $T=40^{\circ}$C; ($\bullet$) $T=29^{\circ}$C; ($\circ$) $T=21^{\circ}$C; 
({\tiny $\blacksquare$}) $T=5.5^{\circ}$C; ($\square$) $T=-4^{\circ}$C;  ($\blacktriangle$) $T=-18^{\circ}$C. Velocity is
plotted on a log scale.}
\label{fig:fig3}
\end{figure}

We present results obtained as follows: under a constant normal stress $p=F_{N}/A$ ($p=25$ MPa for all the results reported hereafter), we measure, at a given temperature, the mean shear stress 
$\sigma=F_{T}/A$ as a function of velocity $V$ in {\it steady sliding}. We have done so for six different temperatures ranging from -18 to 40$^{\circ}$C. The results are presented
on Fig. \ref{fig:fig3}.

We first see that the mean shear stress level increases as $T$ decreases from 40 to -18$^{\circ}$C.

\begin{figure}[ht!]
$$
\includegraphics[width=8cm]{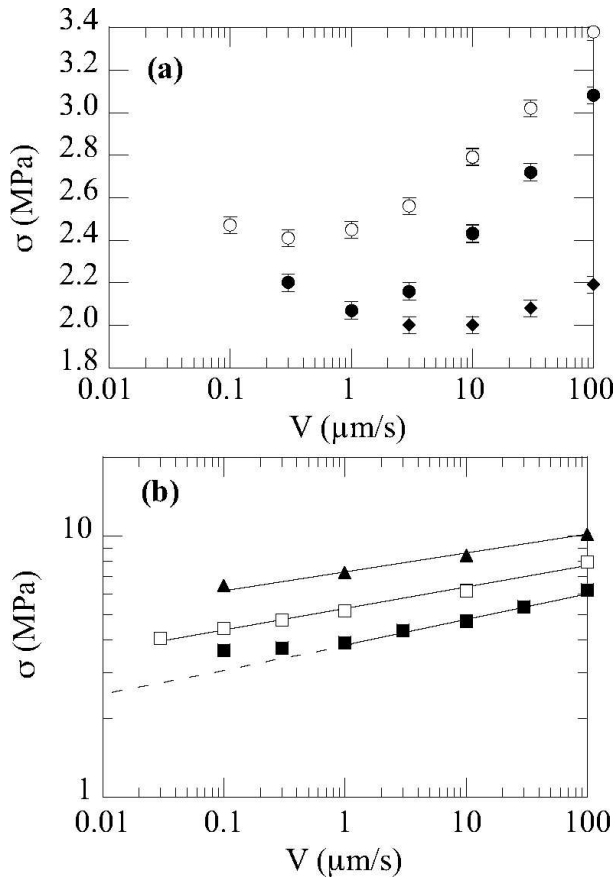}
$$
\caption{Shear stress as a function of velocity, for $p=25$ MPa. (a) ($\blacklozenge$) $T=40^{\circ}$C;  ($\bullet$) $T=29^{\circ}$C; ($\circ$) $T=21^{\circ}$C. 
Velocity is plotted on a log scale. Error bars correspond to the amplitude 
of stress fluctuations observed when sliding at constant $V$, due to large scale (millimeter) chemical heterogeneities of the substrate. (b)
({\tiny $\blacksquare$}) $T=5.5^{\circ}$C; ($\square$) $T=-4^{\circ}$C; ($\blacktriangle$) $T=-18^{\circ}$C . $\sigma$ and $V$
plotted on log scales. Solid lines are power law fits.}
\label{fig:fig4}
\end{figure}

In order to show how the velocity dependence of the shear stress is affected by temperature, the data of Fig. \ref{fig:fig3} have been split into two seperate sets on Fig. \ref{fig:fig4}.
Fig. \ref{fig:fig4}a shows that, at $T=40^{\circ}$C, $\sigma$  increases when V increases from 10 to 100 $\mu$m.s$^{-1}$, and displays a plateau below 10 $\mu$m.s$^{-1}$. Unstable
sliding ({\it i.e.} stick-slip oscillations) is observed at $V<3\, \mu$m.s$^{-1}$. The presence of stick-slip at low $V$ indicates that, below 3 $\mu$m.s$^{-1}$, the shear stress is a decreasing 
function of velocity, which is the source of the sliding instability. The plateau between 3 and 10 $\mu$m.s$^{-1}$ therefore separates a velocity-weakening
from a velocity-strengthening regime, and the shear stress is minimum in this region. 

Such a minimum is clearly visible for
 $T=29^{\circ}$C,  at $V_{c}=1\, \mu$m.s$^{-1}$. A steady velocity-weakening regime ($V<V_{c}$)
is observed for velocities down to $ 0.3\, \mu$m.s$^{-1}$, below which stick-slip sets in.

Decreasing temperature below 29$^{\circ}$C shifts the minimum,
as well as the stick-slip regime, to lower velocities.
At $T=21^{\circ}$C, $V_{c}=0.3\, \mu$m.s$^{-1}$ and stick-slip appears for $V\leq 0.05\mu$m.s$^{-1}$.
At $T=5^{\circ}$C, $\sigma (V)$ only exhibits a plateau on its low velocity side (see Fig. \ref{fig:fig4}b). Stick-slip is observed at $V=0.03\, \mu$m.s$^{-1}$, which signals that $V_{c}$ lies
between 0.03 and 0.1 $\mu$m.s$^{-1}$. Moreover, we note that at high velocities, $\sigma$ increases as $V^{\alpha}$, with $\alpha\simeq 0.1$.
For $T\leq -4^{\circ}$C sliding is steady over the whole range of accessible velocities, and only the $V^{\alpha}$ regime is observed  (see Fig. \ref{fig:fig4}b). The exponent $\alpha$ does
not display any clear sensitivity to temperature.

\begin{figure}[htbp]
$$
\includegraphics[width=8cm]{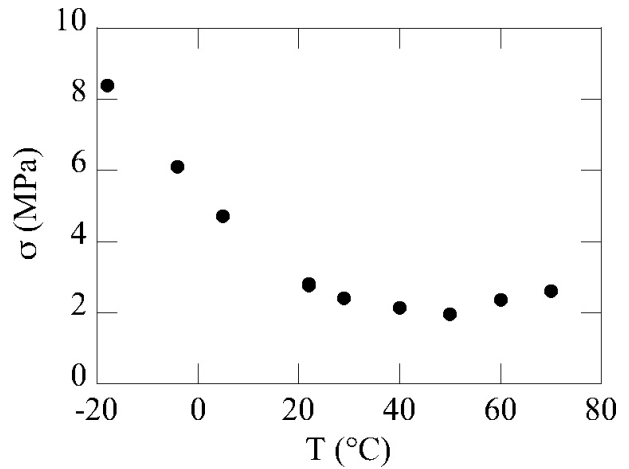}
$$
\caption{Shear stress as a function of temperature at $p=25$ MPa and $V=10\, \mu$m.s$^{-1}$. Error bars are of the size of the symbols.}
\label{fig:fig5}
\end{figure}

In order to better quantify the temperature dependence of the shear stress level, we have performed measurements of $\sigma$ at a fixed sliding velocity  $V_{0}=10\, \mu$m.s$^{-1}$, and extended the
temperature range up to 70$^{\circ}$C. Measurements at $T>70^{\circ}$C were not possible due to bulk creep effects which prevented us from
 maintaining a constant normal stress  during experiments.
$\sigma(V_{0},\,T)$ is plotted on Fig. \ref{fig:fig5}. We see that, in agreement with the above results,
the shear stress displays a fourfold decrease when $T$ varies from $-20$ to $50^{\circ}$C. However, we observe that, at $T=50^{\circ}$C, the trend is reversed, as $\sigma$ 
starts to increase with increasing temperatures above $50^{\circ}$C. 

\section{Discussion}
\label{sec:disc}

Glass transition measurements in supported ultrathin films of poly(methylmethacrylate) indicate that, at the interface with a low energy substrate, the chain mobility is enhanced with respect 
to that of the bulk \cite{nealey}. This agrees with recent positron lifetime measurements which point to the existence of a 2 nm-thick surface layer of lower density \cite{algers}. 
In our experiments, the polymer lens is in contact with a chemically heterogeneous surface (see Fig. \ref{fig:AFM}): the PMMA chains interact with methyl-terminated islands of OTS 
separated by coverage defects. In these defects, which exhibit a much lower silane density, silanol groups (Si-OH) are available to form hydrogen bonds with
the carbonyls of the PMMA sidegroups. The coverage defects thus act as sites on which the polymer chains can get pinned via H-bonding. This recently led us to propose,
in the spirit of Schallamach's model \cite{schal,bo1,revue}, that friction at such an interface is the result of two combined mechanisms \cite{epje2,prl97}:

(i) Viscous dissipation in a thin polymer layer made of chain ends and loops, which yields a velocity-strengthening contribution to the shear stress.

(ii)  Shear induced depinning of surface chains which are adsorbed on the defects. The dynamics of bond formation is governed by a characteristic pinning time $\tau$. Upon
sliding, the faster the chain is driven, the smaller the time available for bond formation. The number of bonds formed at the interface is thus a decreasing function of velocity, which leads to 
a velocity-weakening contribution to $\sigma (V)$. This mechanism is expected to contribute negligibly
to frictional dissipation at sliding velocities above $V_{c}= D/\tau$, where $D$ is an average capture radius ($\sim 10$ nm, the size of a coverage defect).

The position of the minimum in the $\sigma(V)$ curve thus indicates the crossover 
between pinning-controlled and viscosity-controlled friction. Such a picture was found to be consistent with the following observations \cite{prl97}: 
(i) decreasing the size of the pinning sites $D$ 
shifts the crossover to lower velocities, (ii) increasing the contact pressure $p$, which reduces the molecular mobility and increases $\tau$, also lowers $V_{c}$, and (iii) the observed power-law dependence 
of $\sigma$ on $V$, above $V_{c}$, is consistent with the shear-thinning response of strongly confined polymer melts, whose effective viscosity grows with increasing pressure \cite{yamada,robbins}.
This supports the existence of a liquidlike surface layer, as suggested by recent experiments \cite{forrest2}.

The results of the present study give further support to this interpretation.

Fig. \ref{fig:fig4}a shows that the position of the minimum in the $\sigma(V)$ curves depends on temperature: when $T$ increases, $V_{c}$ shifts to higher velocities. This is in agreement with 
the expected decrease of the pinning time $\tau$ when increasing temperature. The presence of a minimum in the $\sigma(V_{0},\, T)$ curve of Fig. \ref{fig:fig5} can be understood as follows. The position of
the minimum in $\sigma(V)$ gradually shifts from below to above the chosen velocity of 10 $\mu$m.s$^{-1}$ as temperature increases from below to above 50$^{\circ}$C. At $T>50^{\circ}$C,
$V_{0}$ enters the pinning-controlled velocity-weakening regime, which leads to the observed stress increase at high
temperature.

For sliding velocities above $V_{c}$, the shear stress increases with velocity as a weak power-law (see Fig. \ref{fig:fig3}c), as already observed in previous experiments. In this regime, the effect of
decreasing the temperature is to shift the $\sigma(V)$ curves to higher shear stress. This is in qualitative
agreement with the fact that, above $V_{c}$, friction results from viscous dissipation in a non-newtonian thin polymer layer confined at the interface, the effective viscosity of which increases when 
temperature decreases. A more quantitative analysis, in order to estimate the effective viscosity of this nanometer-thick 
layer, would require the precise knowledge of the slip boundary condition at the wall, {\it i.e.} the slip-length \cite{leger}, a quantity that cannot be accessed in our experiments.

\begin{figure}[htbp]
$$
\includegraphics[width=8cm]{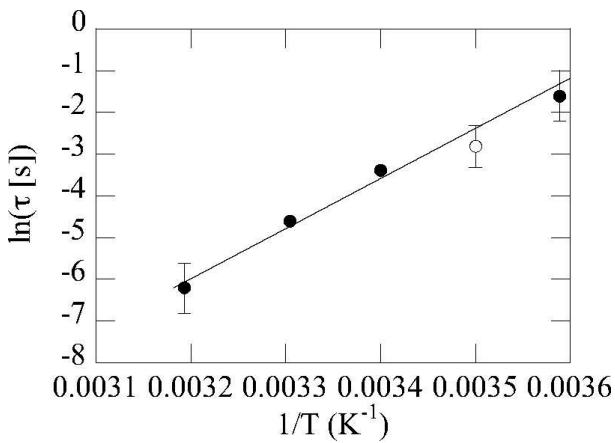}
$$
\caption{Natural logarithm of the pinning time as a function of the inverse of the temperature ($\bullet$). The ($\circ$) symbol corresponds to a value of $\tau_{\beta}$ taken from reference
\cite{rohr}. Error bars on ($\bullet$) symbols correspond to the uncertainty on $V_{c}$ at 5 and 40$^{\circ}$C.
 The line is an arrhenius fit to our the data.}
\label{fig:fig6}
\end{figure}

The existence of a velocity-weakening regime, and of stick-slip oscillations, for $V\leq V_{c}$ is the signature of the interfacial pinning/depinning process.
If we use $V_{c}$ to define a pinning time $\tau=D/V_{c}$, taking $D\simeq 10$ nm, we find:  $\tau$ in the range $10^{-3}$--$3.10^{-3}$ s for 
$T=40^{\circ}$C, $\tau\simeq10^{-2}$ s for 
$T=29^{\circ}$C, 
$3.10^{-2}$ s for $T=21^{\circ}$C, and  $\tau$ in the range $10^{-1}$--$3.10^{-1}$ s at $T=5^{\circ}$C.
Figure \ref{fig:fig6} shows that this pinning time follows an Arrhenius temperature dependence, from which we extract an
activation energy $E_{a}\simeq$20 kcal.mol$^{-1}$. Such an activation energy coincides with that of the $\beta$ relaxation process in PMMA, 
which corresponds to the hindered rotation of the -COOCH$_{3}$ side groups
\cite{hamm,rohr,muzeau}. Furthermore, the values of the pinning time are themselves found to be in good agreement with $\tau_{\beta}$ (see Fig. \ref{fig:fig6}). This leads us to conclude that 
pinning of 
polymer segments on the substrate is governed by $\beta$ rotational motions along the backbone of surface chains.  A pinning dynamics controlled by $\beta$ motions
is indeed consistent with the fact that -C=O groups have to be favorably oriented with respect to silanols in order to form H-bonds. Besides, we note that $\sigma(V)$ exhibits a 
{\it minimum} when $D/V\sim \tau_{\beta}$, from which we conclude that the bulk $\beta$-relaxation, which would give rise to a peak \cite{hamm}, has a negligible contribution to 
frictional dissipation.

\section{Conclusions}

We have shown that macroscopic friction can be a very sensitive probe of the dynamics at the surface
of a glassy polymer. This requires a contact configuration that allows to avoid bulk mechanical losses during sliding. This is achieved, in our experiments, by using a smooth macroscopic lens of polymer
which deforms elastically when it is brought in contact under low pressure with a rigid substrate.  This contrasts with previous studies of glassy polymers using friction force microscopy, where indentation
of the surface by the scanning tip can be such that friction is entirely attributable to bulk dissipation \cite{hamm}. 
Our results are consistent with  the presence of a nanometer-thick layer at the polymer
surface, which, in the case of weak interaction with the countersurface, exhibits a frictional ``rubberlike'' behavior. Frictional dissipation can be ascribed to the combination of (i) viscouslike flow of the
layer, and (ii)  shear induced depinning of polymer chains which can adsorb on defects of the substrate. A quantitative analysis of our data leads us to conclude that the pinning dynamics 
is governed by $\beta$ rotational motions at the interface. 

The author wish to thank Arnaud Arvengas for his contribution to the experiments during his stay at INSP, and 
T. Baumberger and C. Caroli for fruitful comments and discussions.


\begin{thebibliography}{}
\bibitem{note} We have shown, in a previous study, that dissipation occuring in the silane layer contributes negligibly to the measured shear stress \cite{prl97}. Here, we have further checked that 
the observed temperature dependence cannot be ascribed to a
dissipative process in the alkylsilane monolayer. For this purpose, we have replaced the PMMA lens by a clean glass lens. Under comparable nominal contact pressure and at a sliding velocity of 10
$\mu$m.s$^{-1}$, we find that friction does not exhibits any measurable temperature dependence when glass slides on OTS.
\end{thebibliography}
\end{document}